\documentclass[proof]{pasj00}
\newcommand\vecA{\mbox{\boldmath$A$}}
\newcommand\vecB{\mbox{\boldmath$B$}}
\newcommand\vecJ{\mbox{\boldmath$J$}}
\SetRunningHead{K.Otsuji, T.Sakurai, and K.Kuzanyan}{Current Helicity and Twist in Solar Active Regions}
\Received{}
\Accepted{}
\Published{}
\begin{document}
\title{%
Statistical Analysis of Current Helicity and Twist in Solar Active Regions over the Phases of the Solar Cycle Using the Spectro-Polarimeter Data of Hinode
} 
\author{%
Kenichi {\textsc Otsuji}\altaffilmark{1},
Takashi {\textsc Sakurai}\altaffilmark{1,2}, and
Kirill {\textsc Kuzanyan}\altaffilmark{1,3}
}
\altaffiltext{1}{Solar Observatory, National Astronomical Observatory, Mitaka, Tokyo 181-8588}
\altaffiltext{2}{Department of Astronomical Science, The Graduate University for Advanced Studies, Mitaka, Tokyo 181-8588}
\altaffiltext{3}{IZMIRAN, Russian Academy of Sciences, Troitsk, Moscow 142190, Russia}
\email{otsuji@solar.mtk.nao.ac.jp, sakurai@solar.mtk.nao.ac.jp, kuzanyan@izmiran.ru}
\affil{}
\maketitle
\begin{abstract}
Current helicity and twist of solar magnetic fields are important quantities to characterize the dynamo mechanism working in the convection zone of the Sun.
We have carried out a statistical study on the current helicity of solar active regions observed with the Spectro-Polarimeter (SP) of Hinode Solar Optical Telescope (SOT).
We used SOT-SP data of 558 vector magnetograms of a total of 80 active regions obtained from 2006 to 2012.
We have applied spatial smoothing and division of data points into weak and strong field ranges to compare the contributions from different scales and field strengths.
We found that the current helicity follows the so-called hemispheric sign rule when the weak magnetic fields (absolute field strength $< 300$ gauss)
are considered and no smoothing is applied.
On the other hand, the pattern of current helicity fluctuates and violates the hemispheric sign rule when stronger magnetic fields are considered and the smoothing of 2.0 arcsec
(mimicking ground-based observations) is applied.
Furthermore, we found a tendency that the weak and inclined fields better conform to and the strong and vertical fields tend to violate the hemispheric sign rule.
These different properties of helicity through the strong and weak magnetic field components give important clues to understanding the solar dynamo
as well as the mechanism of formation and evolution of solar active regions.
\end{abstract}

\section{Introduction} 
Magnetic or current helicity of solar magnetic fields has become an important and practical measure to assess the dynamics and structure of turbulent motions in the convection zone
and to investigate the driving mechanism of magnetic field generation (the dynamo mechanism) inside the Sun.
Generally speaking, magnetic and cross helicities reflect departure in magnetic and velocity fields from mirror anti-symmetry in a stratified rotating convective envelope.
Magnetic helicity
\begin{equation}
H_{\rm M} = \int \vecA \cdot \vecB \ dV
\end{equation}
cannot be measured directly because the vector potential $\vecA$ ($\vecB = {\rm curl} \vecA$) is involved.
However, electric current helicity
\begin{equation}
H_{\rm C} = \int \vecB \cdot \vecJ \ dV, \qquad \vecJ \equiv {\rm curl}\vecB,
\end{equation}
which expresses the correlation between the magnetic field $\vecB$ and its curl $\vecJ$, is defined in terms of a measurable quantity $\vecB$.
Practically $\vecB$ is only measured at the photospheric level of the Sun (the $z=0$ plane in a local Cartesian coordinate frame).
Therefore, in the following we use
\begin{equation}
H_{{\rm C}z} = \int_{z=0} B_z J_z \ dx dy,
\end{equation}
which we call `current helicity' for simplicity.
$H_{{\rm C}z}$ is proportional to the square of the magnetic flux and inverse size of helical vortices in an observed region,
and traces the magnetic helicity of the large-scale dynamo-generated field. 

Another important quantity which can be computed is twist of field lines represented by $\alpha$,
i.e. the ratio between the vertical component of the electric current and the vertical component of the magnetic field,
\begin{equation}
J_z \simeq \alpha B_z .
\label{eq:curlB}
\end{equation}
This quantity has a measure of the inverse length scale of helical structures.

Helicity in highly conducting plasma in the solar convective zone is well conserved,
so that the surface measurements of helicity proxies may reflect the properties of the deep interior in the Sun.
Thus, such proxies can indicate the conditions of the solar dynamo mechanism,
and the derived information can be used in the dynamo theory as important constrains for its model parameters.
 
\section{Studies in the Past}
The values of the magnetic field components at the solar photosphere can be derived from vector magnetograph data,
which are obtained by instruments of the two classes: Filter-type vector magnetographs and Spectro-polarimeters.
The observations with the former class of instruments are usually less demanding
in terms of the amount of parameters on the data on spectroscopic analysis of polarized light and can be performed much easier and quicker.
However, they have disadvantages in correctly interpreting the so-called Stokes profiles to derive the magnetic field parameters.
The latter instruments provide us with more precise information on the spectral profiles while more demanding in terms of the observation
(longer exposure time and larger amount of output data), which later can be inverted by an optimization process to the magnetic field components.
Many ground-based vector magnetographs, although their performance is limited by atmospheric seeing, use the first class of instruments;
they are the instruments at NASA/MSFC \citep{1983SoPh...88...51W}, NAOJ/Mitaka (SFT, \cite{1995PASJ...47...81S}), and NAOC/Huairou (SMFT, \cite{1986AcASn..27..173A}),
to name a few. The second class of instruments are, e.g., the spectro-polarimeter at the Mees Observatory, University of Hawaii \citep{1985SoPh...97..223M},
and SOLIS/VSM at NSO/Kitt Peak \citep{2003ASPC..307...13K}, as well as extremely high spatial resolution space-born Hinode/SOT Spectro-Polarimeter \citep{2008SoPh..249..167T}.

In the past two decade a serious effort has been carried out to derive electric current helicity and twist of magnetic field from photospheric vector magnetograms.
The first pioneering work of \citet{1990SoPh..125..219S} using just 16 individual active regions has established the so-called hemispheric sign rule for helicity:
the negative/positive sign is dominant in the northern/southern hemisphere, respectively.
This regularity was later re-confirmed by \authorcite{1994ApJ...425L.117P} (\yearcite{1994ApJ...425L.117P,1995ApJ...440L.109P}),
\citet{1998ApJ...496L..43B}, and \citet{2004PASJ...56..831H} in bigger statistical samples of vector magnetograms of solar active regions.
It was noted by \citet{2000JApA...21..303B} as well as \citet{2005PASJ...57..481H} and \citet{2008ApJ...677..719P} that this hemispheric rule
and its strictness may vary with the phase of the solar cycle.
Furthermore, \citet{2010MNRAS.402L..30Z} studied a statistically homogeneous systematic data set for current helicity and twist for two sunspot cycles in 1998-2005 and found that,
while the hemispheric sign rule prevailed in the maximum phase of solar cycle,
at some isolated latitudes in the phases of rise and fall of the cycle there is a prominent tendency of reversal of this rule.
Recently, \citet{2014ApJ...783L...1L} reported the temporal variation of the sign of twist using SDO/HMI obtained from 2010 to 2013.
They found that about 75\% out of sample of 151 active regions obey the hemispheric rule.

Our analysis ultimately aims at scrutinizing the above-mentioned hemispheric rule by using a new high resolution data set
obtained with the Spectro-Polarimeter of the Solar Optical Telescope on board the Hinode satellite (Hinode/SOT-SP, \cite{2008SoPh..249..167T}).
The analysis of these data covering from the end of cycle 23 to the beginning of cycle 24 has already shown
that the validity of this rule varied in the decay and rise phases of the solar cycle \citep{2009ApJ...702L.133T,2011ApJ...733L..27H}.
Other analyses of vector magnetic fields for cycle 23, and the current cycle 24 showed similar controversies (e.g. \cite{2013ApJ...772...52G}).

In order to disentangle the message deciphered in helicity for understanding the mechanism of solar activity cycle, we need a systematic approach \citep{2010MNRAS.402L..30Z}.
This approach would reveal mean quantities which are similar to results of statistical averages over the ensemble of turbulent pulsations in mean field dynamo theory.
Namely, the data must be collected over a large spatial and temporal spread so that the derived quantities are statistically significant.
Therefore, instrumental sensitivity, error propagation, data cadence, and spatial resolution must be considered systematically and mutually balanced.
We may also keep in mind that the nature of systematic errors in the data obtained from ground and space may be different,
and some properties found in the data from the ground-based instruments may not be reproduced for the data from space,
suggesting that the former may be due to atmospheric seeing and may be spurious.

\section{Observation and Data Reduction}
The Hinode/SOT-SP Level-2 data used in this study are available from http://www.csac.hao.ucar.edu.
Using the MERLIN (Milne-Eddington gRid Linear Inversion Network, \cite{2007MmSAI..78..148L}) code,
the magnetic field parameters such as field strength $B$, inclination ($\gamma$) and azimuth ($\phi$) angles,
and magnetic filling factor $f$ were derived from raw spectrum data.
The pixel values of magnetic field components $B_x$, $B_y$, and $B_z$ were computed by
\begin{eqnarray}
B_x &=& f B \sin\gamma \cos\phi, \label{eq:Bx} \\
B_y &=& f B \sin\gamma \sin\phi, \\
B_z &=& f B \cos\gamma. \label{eq:Bz}
\end{eqnarray}
The $z=0$ plane is the photosphere because we selected regions not too far from the disk center.
The $180^\circ$ ambiguity in the azimuth angle was resolved using a combination of the potential field approximation method \citep{1985MPARp.212..312S}
and the non-potential magnetic field calculation method (NPFC, \cite{2005ApJ...629L..69G}).
The former method is less time-consuming, and so we used it to parts of magnetograms with relatively simple magnetic configuration,
and weak field below 300 gauss (G) in the absolute value.
The latter method requires more machine power for disambiguation process; therefore we used it for parts of magnetograms with strongly sheared magnetic field.
For reduction of computation time we set the horizontal resolution down to 1 arcsec, and after the process we remapped the disambiguated data back to its original resolution.

\begin{figure} 
\begin{center}
\FigureFile(85mm,106mm){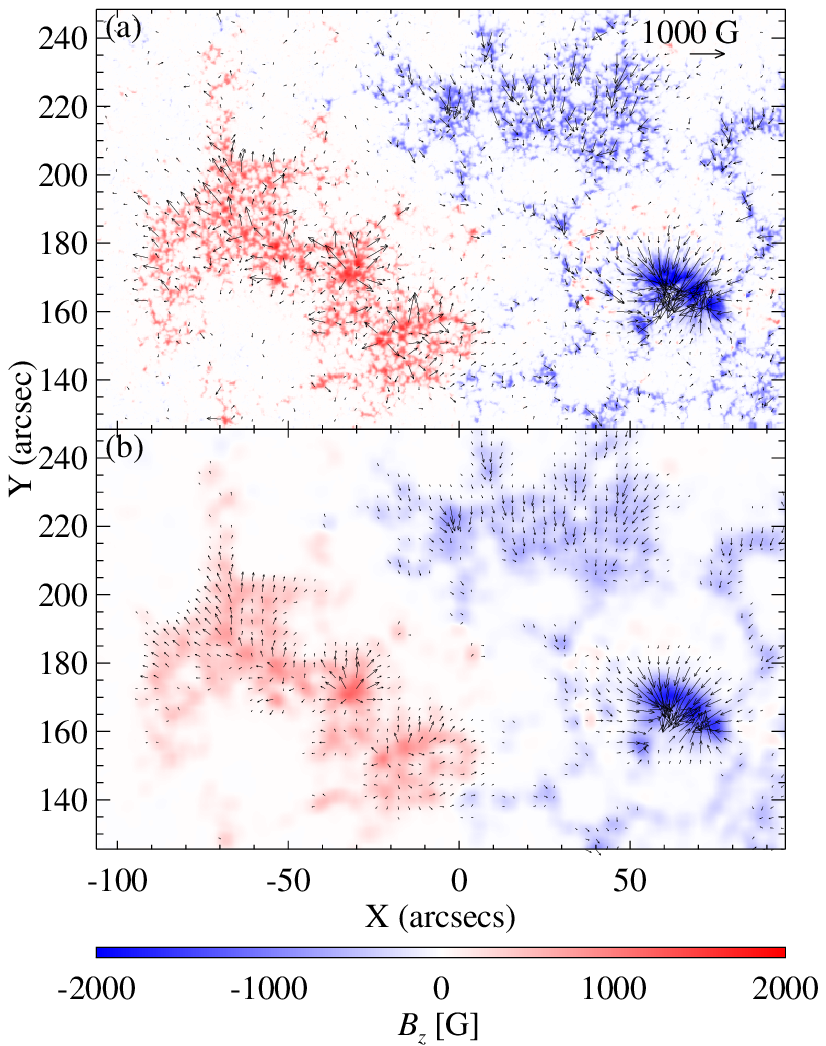}
\caption{Vector magnetic field maps of NOAA 11243 observed on 2011 July 3 from 17:12 UT to 17:55 UT with Hinode/SOT-SP.
The heliographic coordinate of the region is N14$^\circ$ E00$^\circ$.
Red and blue colors represent positive and negative values of $B_z$, and black arrows indicate the horizontal magnetic field.
Panel (a) is without smoothing, while panels (b) is smoothed with 2.0 arcsec Gaussian filter.
Axes (X, Y) are image plane coordinates.}
\label{fig1}
\end{center}
\end{figure}

We used processed Level-2 data obtained in the period from 2006 October 16 to 2012 August 3,
which corresponds to the phase of activity from the end of cycle 23 to the beginning of cycle 24.
We selected magnetograms using the following conditions:
(1) the field-of-view area of SOT/SP is larger than 10,000 arcsec$^2$,
(2) the total area with $|B_z| \geq 1000$ gauss is more than 1,000 arcsec$^2$,
(3) the major part of active region is within the field-of-view of SOT (e.g. if the region is bipolar, both preceding and following main spots are in the field-of-view;
if it is unipolar, the entire main spot is in the field-of-view),
(4) the distance between the active regions and the disk center is less than 30$^\circ$.
We finally obtained in total 558 vector magnetograms of 80 active regions.
Among the selected 80 active regions, 44 are bipolar and 36 are unipolar regions;
63 regions have at least one large sunspot and the remaining 17 regions are pore/plage regions without any prominent sunspot.
The noise levels were determined from these data as 3 gauss for line-of-sight component $B_z$ and 50 gauss for transverse component $B_t = \sqrt{B_x^2 + B_y^2}$.
The data points (pixels) with field components smaller than these thresholds have not been used for computation of the averages over a magnetogram.
Figure \ref{fig1} shows an example of vector magnetic field maps in of a typical active region.
Note that in the following analysis we do not apply transformation to heliographic coordinate system but retain the image plane coordinates $(x,y,z)$ with respect to the observer.
As the distance from the disk center to the active regions is less than 30$^\circ$, in the present study we neglect the projection effect for the magnetic field vector,
and ignore its contribution to statistical averages.
However, we shall study this problem in greater detail in a forth-coming paper.  
Furthermore, as the noise level of $B_t$ is more than 10 times greater than that of $B_z$,
we prefer here to avoid an additional increase in the noise after transformation of the magnetic field vector by mixing different components together.

Then we compute the part of electric current helicity density $H_{{\rm C}z}$ as
\begin{equation}
H_{{\rm C}z} = \left\langle B_z J_z \right\rangle, \qquad
J_z = \frac{dB_y}{dx} - \frac{dB_x}{dy} ,
\end{equation}
where $\langle \cdots \rangle$ represents average over pixels.
In the computation of $J_z$, we mask out the pixels where the disambiguation of horizontal magnetic field is not reliable
(e.g. notable strong discontinuity of $B_x$ or $B_y$), and the pixels where the value of $B_t$ is less than the noise level (50 gauss).
This is to exclude the data from these unreliable pixels from averaging.
We controlled continuity of each of the horizontal field components by limiting the absolute difference between the two neighboring pixels by 350 gauss;
we also limited the absolute difference in the azimuthal angle of the magnetic field vector $\phi$ between the two neighboring pixels by 160$^\circ$.

For the twist $\alpha$, we considered a least-squares procedure \citep{2004PASJ...56..831H},
which determines $\alpha$ in order to minimize the sum of difference between $J_z$ and $\alpha B_z$ using several weighting functions $w(x,y)$.
\begin{equation}
\Sigma w (J_z - \alpha B_z)^2 \Rightarrow {\rm min}
\end{equation}
where $\Sigma$ means summation over magnetogram pixels.
The minimization gives
\begin{equation}
\alpha = \Sigma w B_z J_z \left/ \Sigma w B_z^2 \right.
= \left\langle \alpha(x,y) \frac{w B_z^2}{\langle w B_z^2 \rangle} \right\rangle,
\end{equation}
where $\alpha(x,y) = J_z/B_z$ is a `local' value of twist.
In the following we consider three choices of $w$, namely $wB_z^2 =1, |B_z|$, and $B_z^2$, and correspondingly we designate $\alpha^{(0)}$, $\alpha^{(1)}$, and $\alpha^{(2)}$.
They are averages of $\alpha(x,y)$ with weights $|B_z|^0$, $|B_z|^1$, and $|B_z|^2$, respectively. More explicitly we write
\begin{eqnarray}
\alpha^{(0)} \equiv \alpha_{\rm loc} &=&
\left\langle \frac{J_z}{B_z}\right\rangle, \\
\alpha^{(1)} \equiv \alpha_{\rm av} &=&
\left\langle \frac{J_z \ {\rm sign} (B_z)}{|B_z|}\right\rangle, \\
\alpha^{(2)} \equiv \alpha_{\rm g} &=&
\frac{H_{{\rm C}z}}{\langle B_z^2 \rangle}.
\end{eqnarray}
The average local twist $\alpha_{\rm loc} = \alpha^{(0)}$ \citep{1998ApJ...496L..43B} mainly represents the twist of weak $B_z$ areas
because of $B_z$ in the denominator and because there are numerous pixels in weak-field areas.
The global twist $\alpha_{\rm g} = \alpha^{(2)}$ \citep{2009ApJ...702L.133T}, which is the least-squares fit to the relation $J_z = \alpha B_z$ in a usual sense,
is dominated by (a small number of) pixels with strong $B_z$.
The average twist $\alpha_{\rm av} = \alpha^{(1)}$ \citep{2004PASJ...56..831H} is intermediate between the two.

In this paper we derive these parameters from the original magnetic field data by either applying 0 arcsec (no smoothing), or 2.0 arcsec Gaussian filter.
The 2.0 arcsec smoothing was introduced to mimic typical ground-based observations.

Note that here we have applied Gaussian smoothing on the magnetic field components in order to somehow mimic atmospheric seeing and limited resolution of ground-based telescopes.
However, in real situations the atmospheric effects affect the Stokes polarization spectra.
A more rigorous treatment we may pursue in the future is to apply smoothing to original SP data and reconstruct the values of magnetic field parameters by an inversion algorithm.

\section{Results}
\subsection{Distribution of Local Twist and Helicity within Active Regions}
\begin{figure} 
\begin{center}
\FigureFile(170mm,126mm){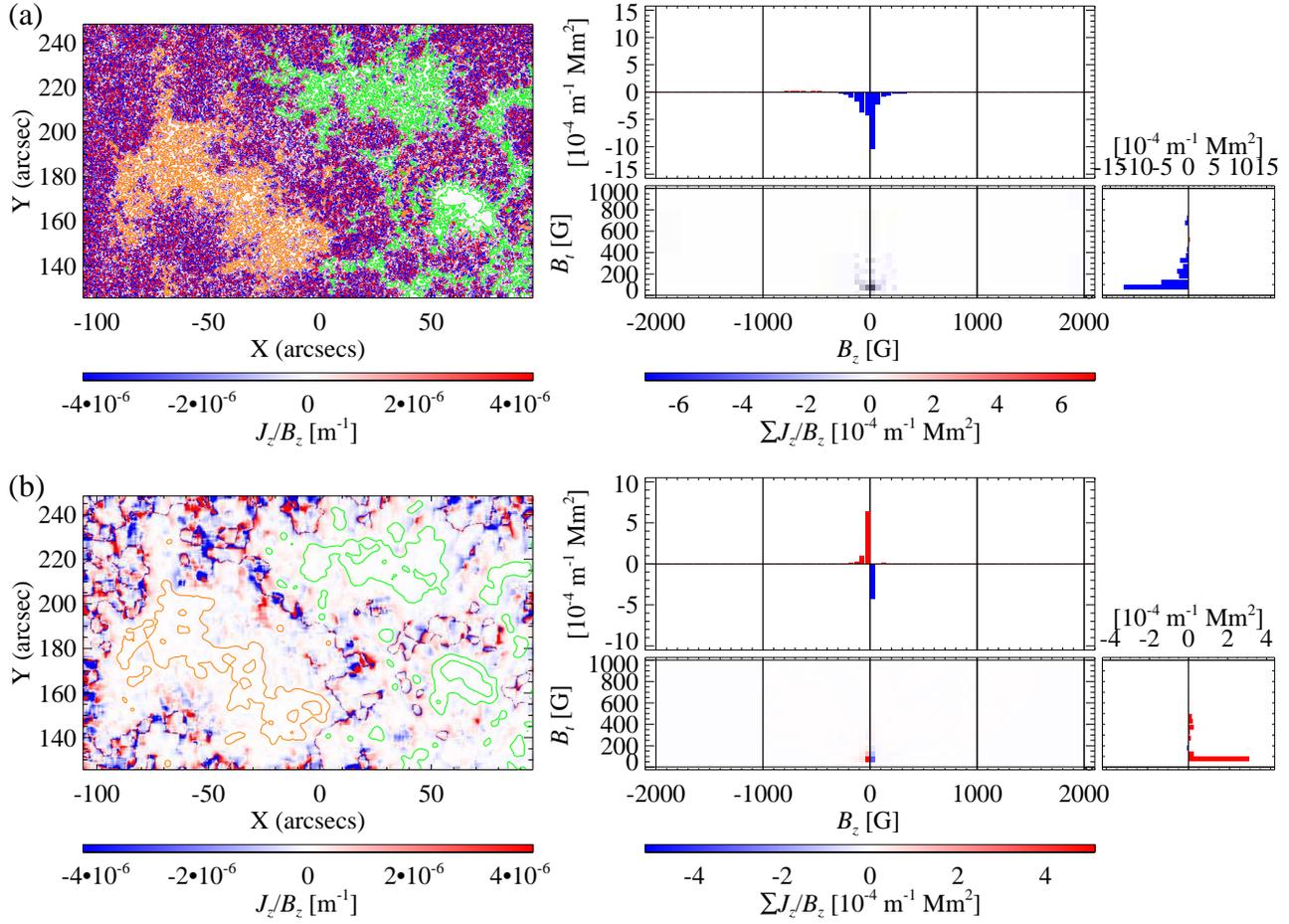}
\caption{The distributions of local twist ($J_z/B_z$) on the spatial $(x, y)$ and the magnetic $(B_z, B_t)$-planes.
(a) and (b) are with no and 2.0 arcsec smoothing, respectively.
Left panels show the maps of local twist $J_z/B_z$.
The orange contours indicate vertical magnetic fields of 300 and 1000 gauss.
The green contours are for negative $B_z$ of the same levels.
The top and right graphs in right panels are typical (one-dimensional) histograms of $J_z/B_z$ against $B_z$ (top) and $B_t$ (right).
The lower left graphs show two-dimensional distributions of $J_z/B_z$ in the $(B_z, B_t)$-plane.
The bin size of both axes is 50 gauss.
The color bar at the bottom indicates the scale for $J_z/B_z$ weighted by the sum of area relative to each magnetic bin.
}
\label{fig2}
\end{center}
\end{figure}

In the following argument we choose the local twist $\alpha(x,y)=J_z/B_z$ and the local current helicity $h_{{\rm C}z}=J_z B_z$ as the magnetic helical parameters to examine
because these two parameters have the different sensitivities to the field strength.
In order to compare the general tendency of how these parameters are distributed in terms of the field strengths, 
we created two-dimensional histograms of $J_z/B_z$ and $J_z B_z$ in the $(B_z, B_t)$-plane for each active region.
For example, figures \ref{fig2}a and \ref{fig2}b show the distributions of $J_z/B_z$ in active region NOAA 11243 observed on 2011 July 3 with/without 2.0 arcsec smoothing.
The left panels of figures \ref{fig2}a and \ref{fig2}b show the spatial distribution of $J_z/B_z$, which concentrates at the magnetically weak field area.
We can see more clearly the distribution of $J_z/B_z$ from the histograms (right panels of figure \ref{fig2}).
The contributions from weak-field regions, which conform to the hemispheric sign rule,
are predominant and almost no contribution from strong-field regions.
This property may explain partly the result that $\alpha_{\rm loc}$ satisfies the hemispheric rule in most cases, as we report in the next section.
Note that the smoothing eliminates strong current $J_z$ by averaging local non-uniformity of horizontal field.
On the other hand, the smoothing increases the vertical field strength at the quiet region just around strong field area. 
As the result, the smoothing decreases the absolute value of $J_z/B_z$.

\begin{figure} 
\begin{center}
\FigureFile(170mm,126mm){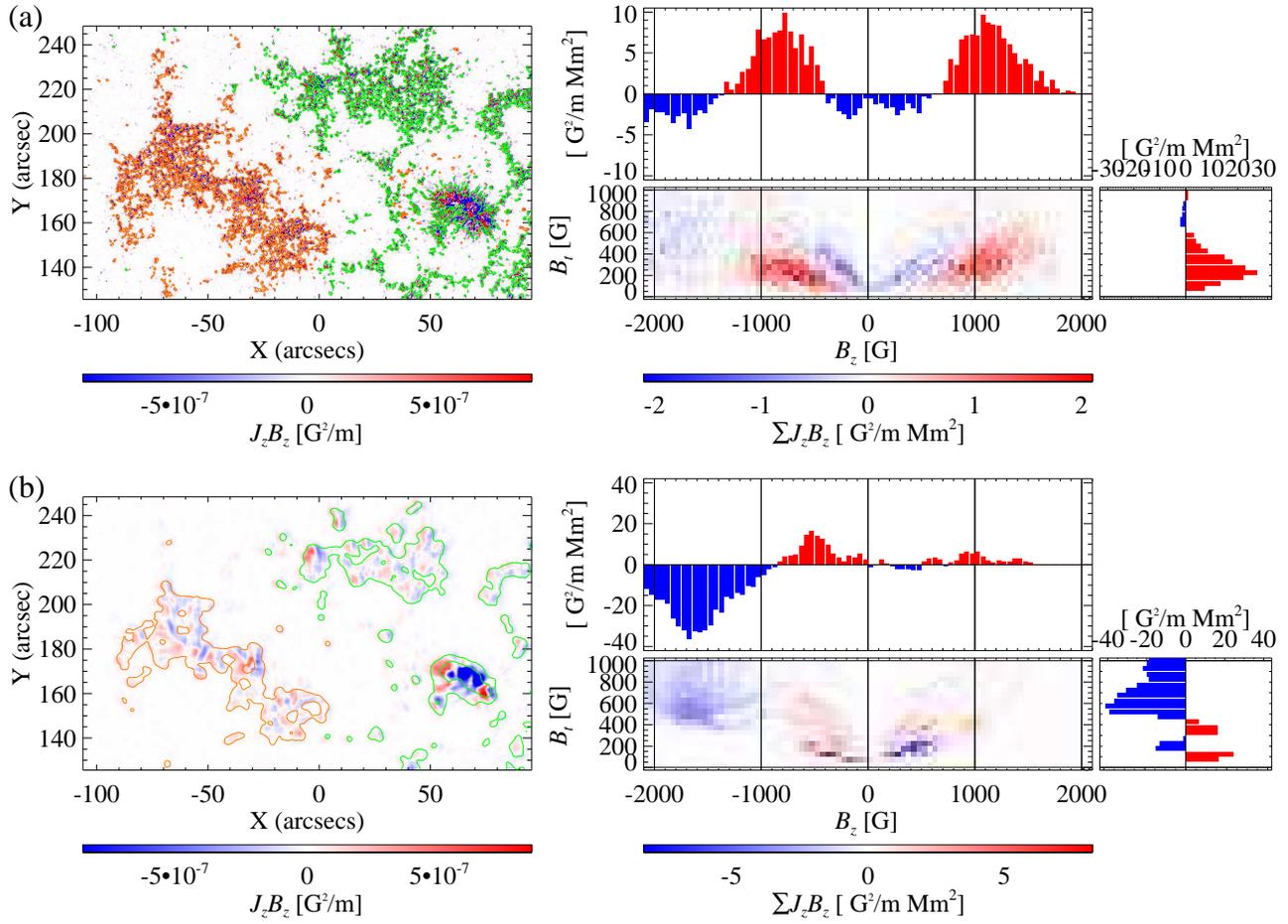}
\caption{The distributions of local current helicity ($J_z B_z$) on the spatial $(x, y)$ and the magnetic $(B_z, B_t)$-planes.
The format is the same as figure \ref{fig2}.}
\label{fig3}
\end{center}
\end{figure}

Figures \ref{fig3}a and \ref{fig3}b show the distributions of local current helicity $J_z B_z$ in the same active region.
We can clearly see that negative helicity is associated with weak ($|B_z|<500$ gauss) and strongly inclined field region while positive helicity is with strong and vertical field.
Considering that this active region was located in the northern hemisphere (N16$^\circ$),
this example suggests that the weak and inclined field conforms to and the strong and vertical field violates the hemispheric sign rule.
This tendency is generally common among developed active regions and plage regions in which the magnetic configuration is relatively simple.
The same tendency was also reported using ground-based observations (\cite{2006ApJ...646L..85Z,2011ApJ...733L..27H}).
On the other hand, we found that active regions with strongly sheared fields such as emerging flux regions show no consistent pattern of helicity distribution
in the $(B_z, B_t)$-plane.
Note that larger smoothing reduces the absolute value of $J_z B_z$.

\subsection{Solar-Cycle Variations in Helicity}
In order to study the solar-cycle variations in helicity and its sign rule, we must specify the time and latitudinal resolution,
and select the data according to this specification.
Since generally more than one magnetogram was taken for one active region, we adopted the following rule.
We assume that the data taken from an active region are tightly correlated when they were taken closely in time
and are more independent when they were taken widely separated in time.
We set the threshold at one day, and for each observing day we average the data to obtain a single value for a helicity parameter that represents the active region. 

We use the latitude bins of $7^\circ$ by following \citet{2010MNRAS.402L..30Z}.
For time bins, we have tried three methods: (i) a fixed time bin of one year, (ii) variable time bins so that each bin contains approximately 30 data sets,
and (iii) plot all the data points at their observing dates. Approach (ii) provides confidence intervals from Student's statistical approach.
We follow earlier applications of statistical considerations of 90\% confidence intervals by \citet{2002ARep...46..424Z}; see also \citet{2010MNRAS.402L..30Z}.
In the following figures we present the most straightforward approach (iii).
In broad terms we can say that the three methods give qualitatively similar results,
but a detailed analysis based on approaches (i) and (ii) is deferred to future papers.


Figure \ref{fig4} shows the butterfly diagrams of $H_{{\rm C}z}$, with/without 2.0 arcsec smoothing and using various magnetic field ranges.
First, all the data points above the noise level ($|B_z|>3$ gauss, $B_t > 50$ gauss) were used to compute these parameters.
Without any smoothing, $H_{{\rm C}z}$ (weighted for larger $|B_z|$) show the opposite pattern (as we can see in figure \ref{fig4}a) of the hemispheric rule,
i.e., the positive/negative sign is dominant in the northern/southern hemisphere, respectively.
We then phenomenologically divided the data points into weak (50 gauss $< |B| <$ 300 gauss),
medium (300 $< |B| < $ 1000 gauss), and strong (1000 gauss $< |B|$) regimes.
When we calculate $H_{{\rm C}z}$ using only the data points of weak magnetic fields, the results follow the hemispheric rule (figures \ref{fig4}b).
On the other hand, $H_{{\rm C}z}$ derived by using only the data points of medium or strong magnetic fields clearly violates the hemispheric rule
(figure \ref{fig4}c for the medium fields\footnote{Strong and medium fields show similar tendencies, but more data points belonging to the medium fields guarantee higher reliability.}).
In all the cases, smoothing on the magnetograms deteriorates the uniformity in the sign of helicity (figures \ref{fig4}d-f).
This can be interpreted as the smoothing merges the medium/strong and weak field areas where the opposite signs are predominant. 

Next, we demonstrate the diagrams for the other parameters, $\alpha_{\rm loc}$, $\alpha_{\rm av}$ and $\alpha_{\rm g}$.
We already pointed out that smoothing mixes the weak and strong field components obeying the opposite hemispheric rules,
which results in the deterioration of the uniformity.
In order to avoid such deterioration, we present only the results derived from no smoothing data (figure \ref{fig5}).
Using all field ranges above the noise level ($|B_z|>3$ gauss, $B_t > 50$ gauss),
$\alpha_{\rm loc}$ (weighted for smaller $|B_z|$) follows the hemispheric rule as we can see in figure \ref{fig5}a,
As the weighting shifts to larger $|B_z|$ (i.e. to $\alpha_{\rm av}$ and $\alpha_{\rm g}$), the hemispheric rule gets reversed as shown in figures \ref{fig5}d and g.
Note that all the $\alpha$ follow the hemispheric rule when we adopt the weak field range of 50 gauss $< |B| <$ 300 gauss (figures \ref{fig5}b, e and h).
In contrast, the butterfly diagrams with the use of medium field range show the opposite tendency to the hemispheric rule (figures \ref{fig5}c, f and i). 


\section{Summary and Discussion}
We found that the current helicity follows the normal hemispheric sign rule when the weak magnetic fields are considered and no smoothing is applied to the Hinode/SOT-SP data.
On the other hand, the pattern of current helicity fluctuates and violates the hemispheric sign rule when stronger magnetic fields are considered and the smoothing
(mimicking image deterioration by atmospheric seeing) is introduced.
Furthermore, we found a tendency that the weak and inclined field better conforms to and the strong and vertical field tends to violate the hemispheric sign rule.
These results show that the so-called hemispheric sign rule is not as simple as it would be if derived from the action of the Coriolis force.
Certainly, the observed tendencies leave opportunity for various interpretations and do not isolate a unique mechanism.

\begin{figure} 
\begin{center}
\FigureFile(170mm,68.7mm){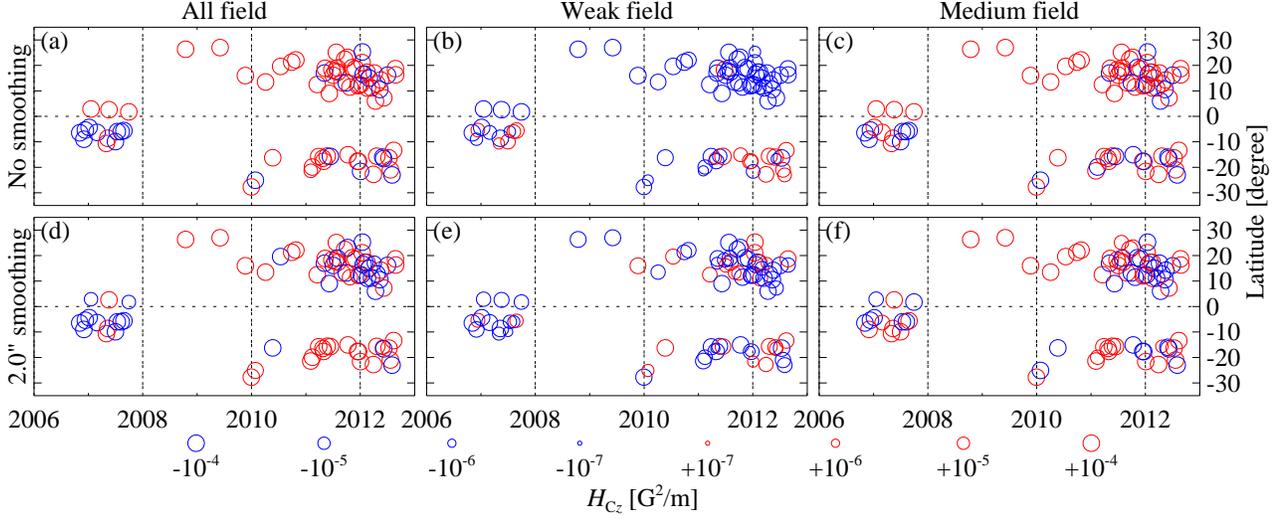}
\caption{%
The butterfly diagrams of $H_{{\rm C}z}$ using various magnetic field ranges.
The left (a and d), middle (b and e) and right (c and f) panels represent the $H_{{\rm C}z}$ distributions on time-latitude planes
corresponding to the magnetic field of all ($|B|>50$ gauss), weak ($|B|=$ 50--300 gauss) and medium ($|B|=$ 300--1000 gauss) ranges, respectively.
The upper (a, b and c) and lower (d, e and f) panels are derived from original (no smoothing) and 2.0 arcsec smoothed data, respectively.  
The time axis at the bottom is years since 2006.
Red (blue) circles indicate active regions which have positive (negative) sign of the parameters.
}
\label{fig4}
\end{center}
\end{figure}

\begin{figure} 
\begin{center}
\FigureFile(170mm,93.9mm){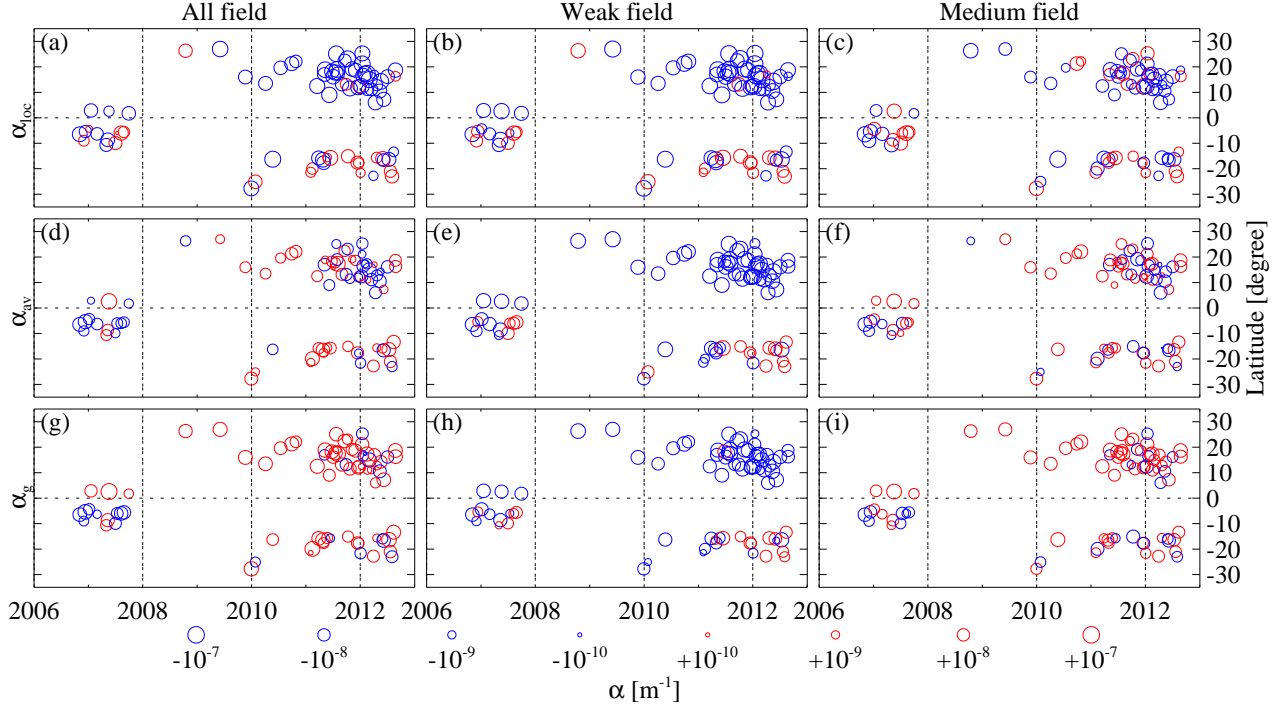}
\caption{%
The butterfly diagrams of $\alpha_{\rm loc}$, $\alpha_{\rm av}$ and $\alpha_{\rm g}$ using various magnetic field ranges without smoothing.
The format is the same as figure \ref{fig4}.
}
\label{fig5}
\end{center}
\end{figure}

We note that figure \ref{fig4} which covers the time span (2006--2012) from the end of solar cycle 23 though the beginning of cycle 24 shows a certain time-latitudinal pattern
that crudely resembles the pattern observed in ground-based observations covering solar cycles 22 and 23 by \citet{2010MNRAS.402L..30Z}.
Namely, they found that the hemispheric helicity law can be regularly inverted to the opposite sign at some latitude in the phases of the solar cycle raise and fall.
While the details vary considerably from one cycle to the other but some qualitative comparison is possible.
We may conclude that while there is no strong support for persistence of that pattern over cycle 24,
there is still no contradiction with the regularity in helicity sign reversals at the beginning and end of the solar cycle
and the validity of the generalized hemispheric sign rule as shown in figure 2 of \citet{2010MNRAS.402L..30Z}.

Any quantity representing the entire helicity consists of parts related to contributions from toroidal and poloidal components of the magnetic field and the electric current.
In the framework of large-scale dynamo models, the part related to the magnetic field is dominated by the toroidal component while the other part related to the electric current is dominated by the poloidal component.
In practical terms it is difficult to separate the contributions from these parts.
The magnetic field as it appears in the solar photosphere is a result of complex magneto-convection and buoyancy processes which mix all parts at a variety of scales.
However, consideration of strong and weak fields enables, very roughly, to separate the influences from weaker (poloidal) component and stronger (toroidal) component.
This fact may explain such a vast difference in the obtained regularities for different ranges of magnetic fields.
Furthermore, other mechanisms of separation of helicity more tightly related to the formation mechanism of sunspots are possible.
Rather different properties of helicity for the magnetic fields of different strength and inclination which we found still require proper theoretical interpretation.

Mimicking the ground-based data by using the Hinode/SOT-SP data is not straightforward.
In addition to spatial smoothing, filter-based magnetographs cannot measure the magnetic filling factor $f$, and instead of equations (\ref{eq:Bx})--(\ref{eq:Bz}), we should use
\begin{eqnarray}
B_x &=& \sqrt{f} B \sin\gamma \cos\phi, \\
B_y &=& \sqrt{f} B \sin\gamma \sin\phi, \\
B_z &=& f B \cos\gamma.
\end{eqnarray}
Using the latter form of computation of the horizontal components of magnetic field we compared the helicity distribution with the results presented in figures \ref{fig4}e and \ref{fig4}f
and found no significant qualitative difference with the former ones.

In addition \citet{2004PASJ...56..831H} applied the selection rule $|B_z| < $ 500 gauss (roughly corresponding to our `weak' field regime) in order to remove the effect of Faraday rotation.
These features must be considered consistently to compare our results with the existing ground-based data sets.

We also feel that some consideration should be given to the unusual behavior of the current solar cycle 24 for which the tendencies noted for previous cycles 22 and 23 may not be relevant.
We hope that future available data from Hinode would help in understanding this tendency.

\bigskip
The authors would like to thank the anonymous Referee for constructive criticism that helped in improvement of the paper.
Hinode is a Japanese mission developed and launched by ISAS/JAXA, with NAOJ as domestic partner and NASA and STFC (UK) as international partners.
It is operated by these agencies in co-operation with ESA and NSC (Norway).
The authors are deeply grateful to the visiting professorship of NAOJ as well as JSPS visiting fellowship programme,
and the staff of NAOJ Solar Physics Division for supporting the visit to NAOJ and the fruitful discussions.



\clearpage

\end{document}